\documentstyle[preprint,prl,aps]{revtex}
\tightenlines

\begin{document}
\draft
\title{Angular momentum sharing in dissipative collisions}

\author{G.~Casini, G.~Poggi, M.~Bini, S.~Calamai, P.~Maurenzig, A.~Olmi,
G.~Pasquali, A. A.~Stefanini, N.~Taccetti}
\address{Istituto Nazionale di Fisica Nucleare and Universit\`a di
         Firenze, I--50125 Florence, Italy}
\author{J. C.~Steckmeyer, R.~Laforest~\cite{byline1}}
\address{Laboratoire de Physique Corpusculaire,
                IN2P3-CNRS, ISMRA et Universit\'e, F-14050
               Caen-Cedex, France}
\author{F.~Saint-Laurent~\cite{byline2}}
\address{GANIL, BP 5027, 14021 Caen-Cedex, France}
\date{\today}
\maketitle
\begin{abstract}
Light charged particles emitted by the projectile-like fragment were
measured in the direct and reverse collision of $^{93}$Nb and
$^{116}$Sn at 25 AMeV.
The experimental multiplicities of Hydrogen and Helium particles as a
function of the primary mass of the emitting fragment show evidence
for a correlation with net mass transfer.
The ratio of Hydrogen and Helium multiplicities points to a dependence
of the angular momentum sharing on the net mass transfer.
\end{abstract}
\narrowtext

It is now experimentally established that binary or quasi-binary
dissipative processes continue to dominate the heavy-ion reaction
cross-section well into the intermediate energy regime
\cite{CharityMo1:91,StefMo2:95,Lott:92,Beau:96,Borde:97,Laroch:98,INDRA}.
In this context, a still open field of investigation concerns the
degree of equilibrium attained in the internal degrees of freedom and,
in particular, the partition of energy~\cite{TokeRev:92} and angular
momentum between the two reaction partners. 

At low bombarding energies ($\leq$15 AMeV), several experimental
findings (concerning mass and charge drift, variances, excitation
energies of reaction products) are rather well accounted for, in some
cases also quantitatively, by models based on the stochastic exchange
of single nucleons (see, e.g., \cite{Rand:82}). 
At larger bombarding energies, the relevance of such a mechanism
becomes somewhat uncertain, due to the decrease of interaction times, 
to the increasing importance of the reaction dynamics and to
associated non-equilibrium effects. 

In the $^{120}$Sn + $^{100}$Mo collision at 19.1 AMeV the fission
probability $P_{fiss}$ of the projectile- and target-like fragment
(PLF and TLF) was measured~\cite{CasiniPfis:91} as a function of the
primary mass $A$.
For a given $A$, corresponding to different net mass transfers for PLF
and TLF, $P_{fiss}$ was found to be significantly larger for the TLF
(which gained mass), even at large TKEL (Total Kinetic Energy Loss).
The observed effect is a clear signature of the lack of an overall
equilibrium between the two partners at the end of the interaction. 

In the $^{100}$Mo + $^{120}$Sn collision at 14.1 AMeV~\cite{Casini:97}
a similar behavior was found also in the binary exit channel, where the
highly excited fragments de-excite mainly by light particle emission.
The observed correlation between the total number of emitted nucleons
and the net mass transfer indicates a non-equilibrium excitation energy
partition between the reaction products, with an excess of excitation
being deposited in the fragment which gains nucleons. 
Similar conclusions had been drawn by other
authors~\cite{Benton:88,Wilcz:89,Kwiat:90,TokePRC:91,Fiore:94},  
but remained quite controversial. 
Although the existence of such correlations is compatible, by itself,
with a nucleon exchange picture, the fact that they are largely
independent of the degree of inelasticity~\cite{Casini:97,TokePRC:91} is
difficult to understand within the present versions of the stochastic
nucleon exchange model and deserves new investigation. 

Up to now no study has been performed concerning possible correlations
between net mass transfer and angular momentum sharing.
This letter presents for the first time a direct evidence for a 
correlation between angular momentum sharing and net mass transfer.

Beams of $^{93}$Nb and $^{116}$Sn at 24.9~AMeV were delivered by the 
GANIL accelerator with an excellent time resolution (about 550~ps and
350~ps FWHM for Nb and Sn, respectively).
The system $^{93}$Nb + $^{116}$Sn was studied in direct and reverse
kinematics, as already done in a previous experiment at lower
energy~\cite{Casini:97}.  
This method allows to obtain information on both partners of the
collision, although the detection is optimized for the PLF and its
associated particles. 
It is equivalent to studying, in two separate runs, the PLF
and TLF of the Nb+Sn collision.
The slight mass asymmetry guarantees a common range of PLF masses in
the exit channels in both kinematic cases, even at moderate TKEL. 
Beam and target feasibility reasons (as well as the wish to have a
system near to the $^{100}$Mo+$^{120}$Sn studied at 14
AMeV~\cite{Casini:97}) guided the choice of the system.
The isotope of Sn, while abundant enough to make a beam, has an
N/Z ratio similar enough to that of $^{93}$Nb so as to reduce the
possible role of isospin.

Heavy ($A\geq20$) reaction products were detected with
position-sensitive gas detectors~\cite{CharityMo1:91,StefMo2:95}.
The FWHM resolution was 3.5 mm for position and 600-750~ps for the
time-of-flight (including the beam contribution).
From the measured velocity vectors, primary (pre-evaporative)
quantities were deduced event-by-event with an improved
version~\cite{CasiniNim:89} of the kinematic coincidence method. 
For elastic events, the FWHM resolution of the primary mass was
$\approx$~2 amu.
The background of incompletely measured events of higher multiplicity
was estimated and subtracted from the results~\cite{CasiniNim:89}.
Behind the forward gas detectors, 
on one side of the beam, an array of 46 Silicon detectors
(covering a sizeable region below and around the grazing angle)
allowed to deduce the secondary (post-evaporative) mass of the
PLF~\cite{Casini:99}.  
Light charged particles were measured with the scintillator array
``Le~Mur''~\cite{refMUR}.
It consists of 96 pads of fast plastic scintillator NE102, 2 mm thick
(threshold $\approx$ 3.2 AMeV for protons and $\alpha$-particles) and
covered in an axially symmetric geometry the region from 2$^{\circ}$
to 18.5$^{\circ}$ behind the gas and Silicon detectors.  
However, because of the shadows of these detectors, not all
scintillator pads could be used.   
More details on the experimental set-up and analysis method are
given in Ref.~\cite{Casini:99}.

The data presented in this letter are focused on binary events in
which light charged particles were detected in the scintillators
in coincidence with PLF and TLF, additionally requiring that the
PLF hits a forward gas detector and one of the Silicon detectors
behind it. 

Results about the analysis of the average number $\Delta A$ of
nucleons emitted by the PLF as a function of its primary mass $A$
are presented in detail elsewhere~\cite{Casini:99}. 
Here we just need to mention that $\Delta A$ versus $A$ gives origin
to two distinct correlations for the Nb+Sn and Sn+Nb data, similarly
to what already observed in the system $^{100}$Mo + $^{120}$Sn at 14
AMeV~\cite{Casini:97}, which gave a direct evidence for the dependence
of $\Delta A$ on the net mass transfer. 

The data of ``Le Mur'' were used to deduce the multiplicity of light
charged particles emitted by the PLF source. 
The limited solid angle of the scintillator array already selected
mainly light particles emitted by the PLF and this geometric selection
was further strengthened in the analysis by rejecting all slow
particles stopped in the pads. 
The remaining particles were cleanly identified in charge Z but
their isotopic composition could not be determined. Therefore in the
following we will just refer to Hydrogen and Helium ions.

Because of the shadows produced by the Silicon detectors,
only light particles emitted on the other side of the beam with
respect to the PLF were considered in the analysis.
Moreover, for a cleaner selection of the PLF source, only light
particles emitted in a forward range (from about 14$^{\circ}$ to
70$^{\circ}$) in the PLF frame were considered. 
In this frame, the experimental light-particle velocity spectra
display an evaporation-like shape, consistent --- within
experimental errors --- with the results of evaporation calculations
obtained with the statistical code GEMINI~\cite{CharityGEM}. 
To deduce the multiplicities of particles emitted by the PLF,
Monte Carlo efficiency corrections were performed, assuming an
evaporative isotropic emission.
Due to these corrections
the absolute values of the light particle multiplicities may be
affected by uncertainties up to $\pm$30\%.
However, the uncertainty is much smaller (of the order of $\pm$10\%)
on the relative values.

The average multiplicities $\langle M_{H} \rangle$ and 
$\langle M_{He} \rangle$ of Hydrogen and Helium ions were determined
as a function of the primary mass $A$ of the emitting PLF,
for bins of TKEL.
The left and right columns of Fig.\ \ref{f:Mlcp-A} present the
results for Hydrogen and Helium, respectively.
The circles and squares refer to the light charged particles emitted
from the PLF in the $^{93}$Nb + $^{116}$Sn and $^{116}$Sn + $^{93}$Nb
reaction, respectively. 
Two assumptions on the excitation energy sharing were used in the
Monte Carlo simulations to deduce the absolute multiplicities from the
experimental ones.
The solid symbols show the results obtained under the assumption of a
non-equilibrium excitation energy sharing (dependent on the net mass
transfer), as deduced from the analysis of the correlations between
$\Delta A$ and $A$~\cite{Casini:99}. 
The open symbols are the results obtained from the same data assuming
an energy sharing independent of the net mass transfer. 
It is apparent that the results are rather insensitive to the particular
physical hypothesis on energy sharing used for the correction. 
The absolute multiplicities for the channel without net mass transfer
are lower than the values obtained by GEMINI calculations by a factor
of about 1.5--2.  
However, they are in fair agreement with the experimental values
obtained for the PLF emission in the similar systems Mo + Mo at 23.7
AMeV~\cite{Gnirs:91} and Xe + Sn at 25 AMeV~\cite{Genouin:99}.

Concerning the dependence on the primary mass $A$ of the PLF,
the multiplicities grow with increasing $A$.
We want to draw attention on the fact that, for a given TKEL,
the multiplicities in the symmetric exit channel are quite different
in the two kinematic cases, although the mass of the emitting PLF is
the same ($A \approx$ 105).
They are larger in the direct reaction (where the PLF has gained mass)
with respect to the reverse reaction (where it has lost an equal
amount of nucleons). 
Being the multiplicity of light charged particles an increasing
function of the excitation energy of the emitter, one can deduce that
the nucleus gaining mass is more excited than the one loosing mass.
This behavior is evident beyond errors at all TKEL values for the
Helium data, whereas for Hydrogen it becomes weaker with increasing TKEL.

These results on the light charged particles are thus qualitatively in
good agreement with those concerning the total number of evaporated
nucleons, $\Delta A$, deduced on the base of a kinematic reconstruction. 
It is worth noting that the measurements of light charged particles in
the scintillator array ``Le Mur'' and of heavy reaction products in
the gas detectors are independent. 
Therefore the results of Fig.\ \ref{f:Mlcp-A} should be little affected
by ``instrumental'' correlations of the kind discussed in
Ref.~\cite{TokeNim:90}.  

As the relative multiplicities are less uncertain than the absolute
ones, we further concentrated on the multiplicity ratio of
Hydrogen and Helium, $\langle M_H \rangle/\langle M_{He} \rangle$. 
Figure\ \ref{f:hhegem}(a) shows this ratio for the exit
channel without net mass transfer.
The solid dots refer to the direct reaction $^{93}$Nb + $^{116}$Sn at
25~AMeV, the solid squares to the reverse reaction at the same
bombarding energy. 
Based on the results for the average excitation energy
partition~\cite{Casini:99}, the excitation energy of the PLF is
estimated from TKEL assuming an equal division of the total excitation
energy, although all arguments that follow are rather insensitive to
this hypothesis.   
The two sets of experimental data are very similar, as expected from
the weak dependence of the light particle multiplicities on the mass
of the emitter. 
In both cases the ratio $\langle M_H \rangle/\langle M_{He} \rangle$
strongly decreases with increasing excitation energy.

It is known since long that in a statistical decay process of a
moderately excited nucleus the ratio of Hydrogen to Helium decreases
with increasing spin of the emitter~\cite{Thomas:64,Gilat:71,Catchen:80}.
Statistical-model calculations with GEMINI~\cite{CharityGEM}, suggest
that this finding may still hold in our range of energies and masses.
In Fig.\ \ref{f:hhegem}(a) the open symbols joined by dotted
lines are the results of GEMINI calculations for an excited $^{93}$Nb
at various values of the spin.
Indeed the ratio between Hydrogen and Helium particles is found to be
rather sensitive to the angular momentum of the evaporating nucleus, 
with large angular momenta favoring the emission of the more massive
Helium particles with respect to the lighter Hydrogen isotopes. 
The comparison of the experimental data with the calculations suggests
that the transfer of angular momentum from the orbital motion to
the internal degrees of freedom of the colliding nuclei is weak
for low TKEL (peripheral collisions) and becomes larger when going to
larger TKEL values (that is, to more central collisions).
However, quantitative estimates of the spin values are difficult as
there are experimental indications that at large excitations GEMINI
tends to underestimate the emission of intermediate mass 
fragments~\cite{Casini:99}. 

Let us concentrate for the remaining part of this letter on the 
events leading to nearly symmetric mass division ($A \approx$ 105) 
in the exit channel.
Figure\ \ref{f:hhegem}(b) presents the ratio of light charged particles
$\langle M_H \rangle/\langle M_{He} \rangle$ emitted by the PLF
measured in the direct and reverse kinematics for the collision 
$^{93}$Nb + $^{116}$Sn at 25~AMeV (solid dots and squares, respectively).
Here the excitation energy $E^{\ast}$ of the PLF has been estimated
from the measured TKEL assuming an excitation energy division in
agreement with that deduced from the total number of nucleons
evaporated from the fragments, but again
the arguments that follow do not depend on this hypothesis. 

For the PLF measured in the direct reaction ---which therefore
experienced a net mass gain--- the experimental results (dots)
correspond to 
rather low ratios $\langle M_H \rangle/\langle M_{He} \rangle$, thus
being an indication of high spin. 
The opposite holds for the PLF measured in the reverse reaction, which
experienced a net loss of nucleons.  
In this second case the results (squares) correspond to 
larger ratios $\langle M_H \rangle/\langle M_{He} \rangle$, thus
pointing to lower spin of the emitter.
This observation clearly indicates that the net gain of nucleons is 
correlated with an excess not only of excitation
energy~\cite{Casini:97,TokePRC:91,Casini:99}, but also of angular momentum.
The same conclusion would hold true also in case of a different
excitation energy partition, as a relative shift of the two sets of
experimental points along the horizontal axis would leave the
data for the direct reaction below those for the reverse one, thus
indicating in any case an asymmetric sharing of angular momentum.

As verified with GEMINI calculations, 
this interpretation of the data in terms of net-mass-transfer 
dependence of both excitation energy and angular momentum partition
can explain the above mentioned weaker dependence of the Hydrogen
multiplicity on the net-mass-transfer at larger excitations.
In fact, in the direct reaction the simultaneous increase of both
excitation energy and spin gives origin to two opposite effects.
The higher excitation energy of PLF tends to increase the average
multiplicity of Hydrogen, but the larger spin tends to depress it.
On the contrary, for Helium both the higher excitation energy and
the larger spin contribute to the increase of the average multiplicity.

It is worth stressing that although the specific interpretation in
terms of angular momentum sharing requires that the observed light
charged particles emission is of statistical origin, 
in any case the experimental observation proves by itself that no
full equilibrium in the relative degrees of freedom between the two
final nuclei, both with $A\approx$105, has been achieved at the end of
the interaction phase. 
In spite of their equal masses, the two nuclei bear memory of the
different ways they have been produced, either by net gain or loss of
nucleons. 

The present data, together with other observations --- like the
persistence of steep correlations between $\Delta A$ and $A$ even at
large TKEL and the systematically larger than expected mass variances
$\sigma^2_A$ at large TKEL~\cite{Casini:97,Casini:99} --- suggest the
presence of some other mechanism besides the mere stochastic exchange
of single nucleons across a window in the dinuclear system. 

For example,
the fluctuations in the rupture of a long stretched neck might
be an essential ingredient to explain the experimental features.
They can explain in a natural way the observation of huge mass
variances in spite of rather short interaction times. 
Asymmetric neck ruptures, with the lighter collision partner
obtaining a larger share of the neck matter, could sizably contribute 
to the net mass transfer leading to symmetric mass division \cite{Laroch:98}.
In this case the two final nuclei, although of equal mass, would
strongly differ in shape and moment of inertia. 
This neck remnant, if re-absorbed, might be responsible for the larger
share of angular momentum. 
Or it might contribute to the increased production of intermediate
mass fragments in the region between the two collision partners, which
is presently a highly debated topic.

Concluding, the collision $^{93}$Nb + $^{116}$Sn at 25 AMeV has been
studied in direct and reverse kinematics.
The light charged particles emitted by the PLF allow a qualitative
estimation of the transferred angular momentum and give evidence of a 
situation of non-equilibrium existing between the two collision
partners at the end of the interaction.
In fact the net gain of nucleons appears to be correlated with an
excess of both excitation energy \underline{\it and} angular momentum. 
This experimental finding seems difficult to reconcile with existing
models based on stochastic exchanges of singles nucleons and calls for
a better theoretical understanding of the microscopic interaction
mechanism, including other effects like, e.g., an explicit
treatment of the neck degrees of freedom.

We wish to thank the GANIL staff for delivering high quality beams
pulsed with very good time structure.  
We also thank R. Ciaranfi and M. Montecchi for their skillfulness in
the development of dedicated electronic modules and P. Del Carmine and
F. Maletta for their valuable support in the preparation of the
experimental set-up.

\begin{figure}
\caption{Efficiency corrected experimental multiplicities for Hydrogen
(left column) and Helium particles (right column) emitted by the PLF, 
plotted as a function of the primary mass $A$ of the PLF, for
100 MeV wide bins of TKEL.  
The circles and squares refer to light charged particles emitted from
the PLF in the $^{93}$Nb + $^{116}$Sn and $^{116}$Sn + $^{93}$Nb
reaction, respectively. 
The solid and open symbols show the experimental data corrected
assuming an excitation energy sharing dependent on net mass transfer
(as deduced from $\Delta A$ of the fragments) and independent
of it, respectively.
The lines are guides to the eye.
The error bars show the statistical errors.}
\label{f:Mlcp-A}
\end{figure}

\begin{figure}
\caption{(a) Ratio of the experimental average multiplicities of
Hydrogen and Helium particles, 
$\langle M_H \rangle/\langle M_{He} \rangle$, emitted by
the PLF in exit channels without net mass transfer. 
The solid dots (squares) refer to PLF from the direct (reverse)
collision $^{93}$Nb + $^{116}$Sn at 25~AMeV and are plotted as a
function of the excitation energy estimated from the data.
For comparison also the results of GEMINI calculations for several
values of the spin of an emitting $^{93}$Nb-source are shown. 
The lines are drawn to guide the eye.
(b) Same presentation as in part (a) except that the data
refer to events leading to a symmetric mass division ($A\approx$105)
in the exit channel.}
\label{f:hhegem}
\end{figure}


\begin{thebibliography}{10}

\bibitem[*]{byline1}
 Permanent Address: 
Washington University, Medical School, BOX 8225
510 Kingshiway
St-Louis, MO 63110


\bibitem[\dagger]{byline2}
 Permanent Address:
DRFC/STEP, CEA/Cadarache, F-13108 Saint-Paul-lez-Durance Cedex.

\bibitem{CharityMo1:91}
 R. J. Charity {\it et~al.}, Z.\ Phys.\ A {\bf 341}, 53 (1991).

\bibitem{StefMo2:95}
 A. A. Stefanini {\it et al.}, Z.\ Phys.\ A {\bf 351}, 167 (1995).

\bibitem{Lott:92}
 B. Lott {\it et~al.}  Phys. Rev. Lett. .{\bf 68}, 3141 (1992). 

\bibitem{Beau:96}
 L. Beaulieu {\it et~al.}  Phys. Rev. Lett. .{\bf 77}, 462 (1996). 

\bibitem{Borde:97}
 B. Borderie  {\it et~al.}, Z.\ Phys.\ A {\bf 357}, 7 (1997).

\bibitem{Laroch:98}
 Y. Larochelle  {\it et~al.}, Phys.\ Rev.\ C {\bf 57}, R1027 (1998). 

\bibitem{INDRA}
 J. Lukasik {\it et~al.}, Phys.\ Rev.\ C {\bf 55}, 1906 (1997).

\bibitem{TokeRev:92}
 J. T\~{o}ke and W. U. Schr\"oder, 
     Ann.\ Rev.\ Nucl.\ Part.\ Sci.\ {\bf 42}, 401 (1992).

\bibitem{Rand:82}
J. Randrup, Nucl.\ Phys.\ A {\bf 307}, 319 (1978); 
                            {\bf 327}, 490 (1979);
                            {\bf 383}, 468 (1982).

\bibitem{CasiniPfis:91}
 G. Casini {\it et~al.}, Phys.\ Rev.\ Lett. {\bf 67}, 3364 (1991).

\bibitem{Casini:97}
 G. Casini {\it et~al.}, Phys.\ Rev.\ Lett. {\bf 78}, 878(1997).

\bibitem{Benton:88}
 D. R. Benton {\it et al.}, Phys.\ Rev.\ C {\bf 38}, 1207 (1988).

\bibitem{Wilcz:89}
 J. Wilczynski {\it et~al.},  Phys.\ Lett.\ B {\bf 220}, 497 (1989).

\bibitem{Kwiat:90}
 K. Kwiatkowski {\it et~al.}, Phys.\ Rev.\ C {\bf 41}, 958 (1990).

\bibitem{TokePRC:91}
 J. T\~{o}ke {\it et~al.}, Phys.\ Rev.\ C {\bf 44}, 390 (1991).

\bibitem{Fiore:94}
 L. Fiore {\it et~al.}, Phys.\ Rev.\ C {\bf 50}, 1709 (1994).

\bibitem{CasiniNim:89}
 G. Casini {\it et~al.}, Nucl.\ Instr.\ Methods\ A {\bf 277}, 445 (1989).

\bibitem{Casini:99}
 G. Casini {\it et~al.} to be published.

\bibitem{refMUR}
 G. Bizard  {\it et~al.}  Nucl.\ Instr.\ Methods\ A {\bf 244}, 483 (1986).

\bibitem{CharityGEM}
 R. J. Charity {\it et~al.}, Nucl.\ Phys.\ A {\bf 483}, 371 (1988);
                                             {\bf 511},  59 (1990).
\bibitem{Gnirs:91}
 M. Gnirs, PhD thesis, Heidelberg University, 1991.

\bibitem{Genouin:99}
 E. Genouin-Duhamel, PhD thesis, Caen University, 1999.

\bibitem{TokeNim:90}
 J. T\~{o}ke {\it et al.}, Nucl.\ Instr.\ Meth.\ A {\bf 288}, 406 (1990).

\bibitem{Thomas:64}
 T. D. Thomas, Nucl.\ Phys. {\bf 53}, 538 (1964).

\bibitem{Gilat:71}
 J. Gilat and J. R. Gover, Phys.\ Rev.\ C {\bf 3}, 734 (1971).

\bibitem{Catchen:80}
 G. L. Catchen {\it et~al.}, Phys.\ Rev.\ C {\bf 21}, 940 (1980).


\end{thebibliography}
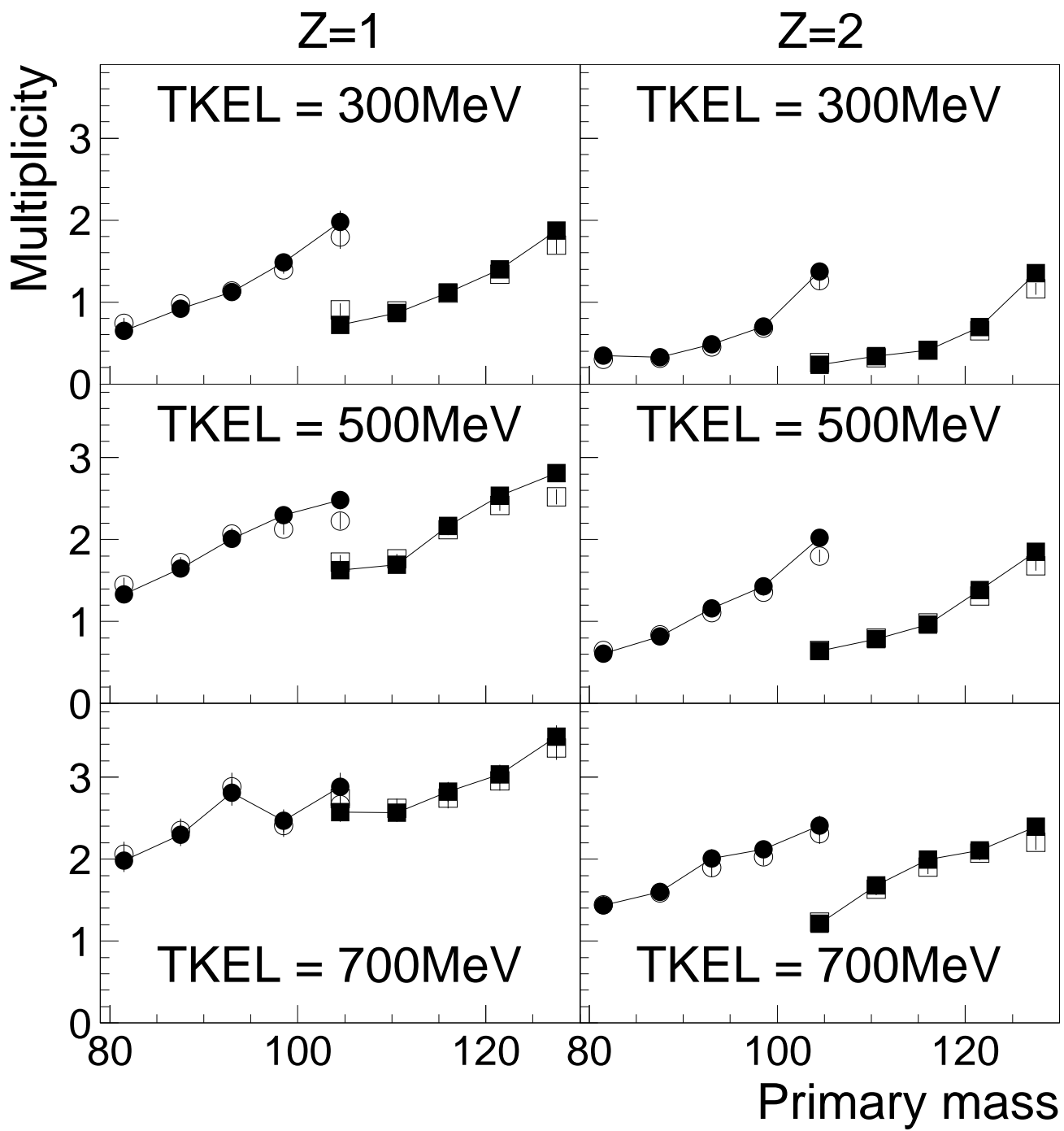
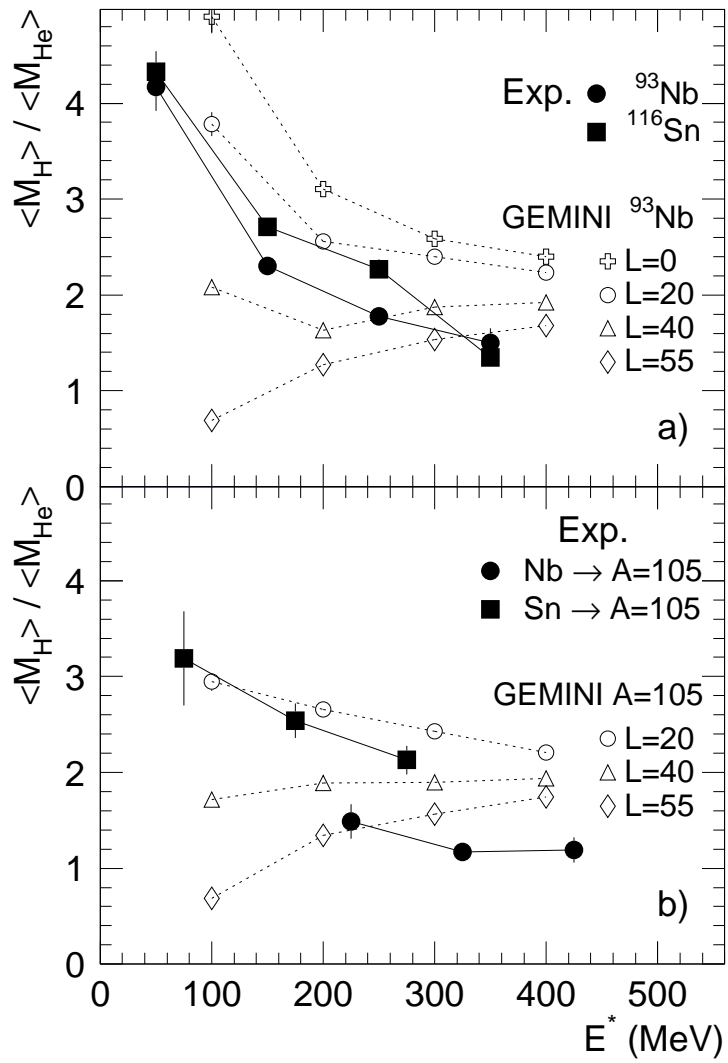
\end{document}